%% file: main.tex
\documentclass[journal,comsoc]{IEEEtran}
\bstctlcite{IEEEexample:BSTcontrol} 

\input{zz_packages}
\newif\ifdraft
\drafttrue

\makenoidxglossaries
\loadglsentries{acronyms}

\input{zz_machinery}
\input{zz_macro}

\providecommand{\ie}{i.e.}
\providecommand{\eg}{e.g.}
\providecommand{\viz}{viz.}

\begin{document}
\bstctlcite{IEEEexample:BSTcontrol}

\title{Operationalizing Quantum Machine Learning for IoT Network Intrusion Detection}

\author{\IEEEauthorblockN{
Domenico Ciuonzo,
Idio Guarino,
Alfredo Nascita,
Antonio Pescapè,
Vincenzo Spadari
}
}

\maketitle

\begin{abstract}
\input{tex/00_abstract.tex}
\end{abstract}

\section*{Introduction}
\label{sec:introduction}
\glsresetall

\input{tex/010_introduction.tex}

\section*{The Advent of Quantum-Enhanced Network Intrusion Detection}
\label{sec:back_rw}
\input{tex/020_related_work}

\section*{Building the Quantum NIDS}
\label{sec:methodology}
\input{tex/030_methodology}

\section*{Experimental Setup}
\label{sec:experimental_results}

\input{tex/050_experimental_results}

\section*{Conclusion}
\label{sec:conclusion}

\input{tex/060_conclusion}

\bibliographystyle{IEEEtranN}
\begingroup
\footnotesize
\bibliography{IEEEabrv,tc_bibliography}
\endgroup

\input{tex/90_biographies.tex}

\vfill

\end{document}

%% file: zz_packages.tex
\usepackage{amsmath,amsfonts,amssymb}
\usepackage{algorithmic}
\usepackage{algorithm}
\usepackage{array}
\usepackage[caption=false, font=footnotesize, labelformat=empty]{subfig}
\usepackage{textcomp}
\usepackage{stfloats}
\usepackage{verbatim}
\usepackage{graphicx}
\usepackage{booktabs}

\usepackage{threeparttable}
\usepackage{fontawesome5}
\usepackage{pifont}
\usepackage[table]{xcolor}
\usepackage{multirow,multicol}
\usepackage[square,sort,comma,numbers,compress]{natbib} \usepackage[acronym,toc]{glossaries}
\usepackage{tcolorbox}
\usepackage{xspace}
\usepackage{bm}
\usepackage{svg}
\usepackage[inline]{enumitem}
\usepackage{makecell}
\usepackage{todonotes}


%% file: acronyms.tex
\newacronym{sota}{SOTA}{state-of-the-art}
\newacronym{rq}{RQ}{Research Question}
\newacronym{ra}{RA}{Research Answer}
\newacronym{aug}{\texttt{Aug}}{Supervised Data Augmentation}
\newacronym{simcon}{\texttt{SimCon}}{Self-supervised Contrastive Learning}
\newacronym{supcon}{\texttt{SupCon}}{Supervised Contrastive Learning}
\newacronym{cl}{\texttt{CL}}{Contrastive Learning}
\newacronym{genai}{GenAI}{Generative Artificial Intelligence}
\newacronym{traffgen}{NTG}{Network Traffic Generation}
\newacronym{traffclass}{NTC}{Network Traffic Classification}
\newacronym{id}{ID}{Intrusion Detection}
\newacronym{nda}{NDA}{Network Digital Assistance for Documentation \& Operation}
\newacronym{nid}{NID}{Network Intrusion Detection}
\newacronym{nids}{NIDS}{Network Intrusion Detection System}
\newacronym{ide}{IDE}{Implicit Density Estimation}
\newacronym{ede}{EDE}{Explicit Density Estimation}

\newacronym{la}{LA}{Log Analysis}
\newacronym{nla}{NSLA}{Networked System Log Analysis}
\newacronym{lad}{LAD}{Log-based Anomaly Detection}

\newacronym{xai}{XAI}{eXplainable Artificial Intelligence}
\newacronym{dl}{DL}{Deep Learning}
\newacronym{qml}{QML}{Quantum Machine Learning}
\newacronym{ai}{AI}{Artificial Intelligence}
\newacronym{iat}{IAT}{Inter Arrival Time}
\newacronym{tcpwin}{WIN}{TCP Window Size}
\newacronym{pl}{PL}{Packet Length}
\newacronym{ps}{PS}{Packet Size}
\newacronym{ttl}{TTL}{Time To Live}
\newacronym{ts}{TS}{Timestamp}
\newacronym{to}{TO}{Traffic Object}
\newacronym{tp}{TP}{Traffic Prediction}
\newacronym{ac}{AC}{Attack Classification}
\newacronym{ad}{AD}{Anomaly Detection}
\newacronym{nad}{NAD}{Network Anomaly Detection}
\newacronym{tc}{TC}{Traffic Classification}
\newacronym{nta}{NTA}{Network Traffic Analysis}
\newacronym{cnn}{CNN}{Convolutional Neural Network}
\newacronym{qnn}{QNN}{Quantum Neural Network}
\newacronym{qcnn}{QCNN}{Quantum Convolutional Neural Network}
\newacronym{hqnn}{HQNN}{Hybrid Quantum Neural Network}
\newacronym{ml}{ML}{Machine Learning}
\newacronym{ids}{IDS}{Intrusion Detection System}
\newacronym{shap}{SHAP}{SHapley Additive exPlanations}
\newacronym{eda}{EDA}{Exploratory Data Analysis}
\newacronym{ae}{AE}{AutoEncoder}
\newacronym{ece}{ECE}{Expected Calibration Error}
\newacronym{ls}{LS}{Label Smoothing}
\newacronym{fl}{FL}{Focal Loss}
\newacronym{lime}{LIME}{Local Interpretable Model-agnostic Explanations}
\newacronym{qos}{QoS}{Quality of Service}
\newacronym{qoe}{QoE}{Quality of Experience}
\newacronym{dt}{DT}{Decision Tree}
\newacronym{md}{MD}{Misuse Detection}
\newacronym{bmd}{bMD}{Binary Misuse Detection}
\newacronym{mmd}{mMD}{Multi-class Misuse Detection}
\newacronym{can}{CAN}{Controller Area Network}
\newacronym{apt}{APT}{Advanced Persistent Threat}
\newacronym{mtu}{MTU}{Maximum Transmission Unit
}
\newacronym{ran}{RAN}{Radio Access Network}
\newacronym{iot}{IoT}{Internet of Things}
\newacronym{iomt}{IoMT}{Internet of Medical Things}
\newacronym{llm}{LLM}{Large Language Model}
\newacronym{lstm}{LSTM}{Long Short-Term Memory}
\newacronym{tsne}{t-SNE}{t-distributed Stochastic Neighbor Embedding}
\newacronym{dir}{DIR}{Packet Direction}
\newacronym{darpa}{DARPA}{Defense Advanced Research Projects Agency}
\newacronym{nlp}{NLP}{Natural Language Processing}
\newacronym{sni}{SNI}{Server Name Indication}
\newacronym{svm}{SVM}{Support Vector Machine}
\newacronym{lrp}{LRP}{Layer-wise Relevance Propagation}
\newacronym{gradcam}{Grad-CAM}{Gradient-weighted Class Activation Mapping}
\newacronym{ig}{IG}{Integrated Gradients}
\newacronym{tls}{TLS}{Transport Layer Security}
\newacronym{dnn}{DNN}{Deep Neural Network}
\newacronym{som}{SOM}{Self Organizing Map}
\newacronym{knn}{KNN}{K-Nearest Neighbors}
\newacronym{pfi}{PFI}{Permutation Feature Importance}
\newacronym{ice}{ICE}{Individual Conditional Expectation}
\newacronym{pdp}{PDP}{Partial Dependence Plot}
\newacronym{cem}{CEM}{Contrastive Explanation Method}
\newacronym{coin}{COIN}{Contextual Outlier INterpretation}
\newacronym{ciu}{CIU}{Contextual Importance and Utility}
\newacronym{brcg}{BRCG}{Boolean Decision Rules via Column Generation}
\newacronym{cade}{CADE}{Contrastive Autoencoder for Drifting detection and Explanation}
\newacronym{deeplift}{DeepLIFT}{Deep Learning Important FeaTures}
\newacronym{lemna}{LEMNA}{Local Explanation Method using Nonlinear Approximation}
\newacronym{sdn}{SDN}{Software Defined Networking}
\newacronym{fpga}{FPGA}{Field Programmable Gate Array}
\newacronym{nmm}{NMM}{Network Monitoring and Management}
\newacronym{api}{API}{Application Programming Interface}
\newacronym{ui}{UI}{User Interface}
\newacronym{cli}{CLI}{Command Line Interface}
\newacronym{rnn}{RNN}{Recurrent Neural Network}
\newacronym{gru}{GRU}{Gated Recurrent Unit}
\newacronym{mlm}{MLM}{Masked Language Model}
\newacronym{ddpm}{DDPM}{Denoising Diffusion Probabilistic Model}
\newacronym{ssm}{SSM}{State Space Model}
\newacronym{vit}{ViT}{Vision Transformer}

\newacronym{gan}{GAN}{Generative Adversarial Network}
\newacronym{vae}{VAE}{Variational Autoencoder}
\newacronym{gasf}{GASF}{Grammian Angular Summation Field}

\newacronym{jsd}{JSD}{Jensen-Shannon Divergence}
\newacronym{tvd}{TVD}{Total Variation Distance}
\newacronym{sr}{SR}{Success Rate}

\newacronym{rfc}{RFC}{Request for Comment}
\newacronym{cpsa}{CPSA}{Cryptographic Protocol Shapes Analyzer}
\newacronym{nf}{NF}{Network Function}
\newacronym{sml}{SLM}{Small Language Model}
\newacronym{gpp}{3GPP}{Third Generation Partnership Project}
\newacronym{rag}{RAG}{Retrieval-Augmented Generation}

\newacronym{fed}{FED}{Full Encoder-Decoder}
\newacronym{eo}{EO}{Encoder-Only}
\newacronym{do}{DO}{Decoder-Only}
\newacronym{sdp}{SDP}{Sequential Denoising Process}

\newacronym{mlp}{MLP}{MultiLayer Perceptron}

\newacronym{peft}{PEFT}{Parameter-Efficient Fine-Tuning}
\newacronym{ptq}{PTQ}{Post-Training Quantization}
\newacronym{lora}{LoRA}{Low-Rank Adaptation}
\newacronym{gguf}{GGUF}{GPT-Generated Unified Format}
\newacronym{nice}{NICE}{Non-linear Independent Components Estimation}
\newacronym{gpu}{GPU}{Graphics Processing Unit}

\newacronym{crps}{CRPS}{Continuous Ranked Probability Score}
\newacronym{fpr}{FPR}{False Positive Rate}

\newacronym{biflow}{BF}{Bi-direction Flow}
\newacronym{flow}{F}{Flow}
\newacronym{nc}{NC}{Network Conversation}

\newacronym{fgsm}{FGSM}{Fast Gradient Sign Method}

\newacronym{ood}{OOD}{Out-of-Distribution}
\newacronym{iid}{ID}{In-Distribution}

\newacronym{qc}{QC}{Quantum Computing}

\newacronym{qaoa}{QAOA}{Quantum Approximate Optimization Algorithm}
\newacronym{qsvm}{QSVM}{Quantum Support Vector Machine}
\newacronym{hcvqnn}{HCVQNN}{Hybrid Continuous Quantum Neural Network}
\newacronym{vqc}{VQC}{Variational Quantum Circuit}
\newacronym{qkm}{QKM}{Quantum $K$-means}
\newacronym{sel}{SEL}{Strongly Entangling Layers}
\newacronym{wel}{WEL}{Weakly Entangling Layers}
\newacronym{pqc}{VQC}{Parameterized Quantum Circuit}

\newacronym{pca}{PCA}{Principal Component Analysis}

\newacronym{mqtt}{MQTT}{Message Queuing Telemetry Transport}
\newacronym{dos}{DoS}{Denial-of-Service}
\newacronym{ddos}{DDoS}{Distributed Denial-of-Service}

\newacronym{nisq}{NISQ}{Noisy Intermediate-Scale Quantum}
\newacronym{pauc}{pAUC}{partial Area Under Curve}
\newacronym{roc}{ROC}{Receiver Operating Characteristic}

\newacronym{oc}{OC}{Operational Challenge}

%% file: zz_machinery.tex
\usepackage{silence}
\usepackage{comment}

\specialcomment{rimuovere}{}{}
\specialcomment{rileggere}{\begingroup\color{teal}}{\endgroup}
\specialcomment{commentolungo}{\begingroup\color{gray}}{\endgroup}
\specialcomment{new}{\begingroup\color{blue}}{\endgroup}
\specialcomment{warning}{\begingroup\color{red}}{\endgroup}

\ifdraft
    \usepackage[]{hyperref} 
                     
          \usepackage{todonotes}

        \specialcomment{rewrite}{\begingroup\color{red}}{\endgroup}
    \specialcomment{draft}{\begingroup\color{blue}}{\endgroup}

\else   
        \usepackage[draft]{hyperref}     

             \excludecomment{rewrite}
    \excludecomment{draft}

    \excludecomment{daicc}
\fi

%% file: zz_macro.tex
\newcommand{\ciciomt}{\textsc{CIC-IoMT2024}\xspace}
\newcommand{\edge}{\textsc{Edge-IIoT}\xspace}

\newcommand{\iotnidd}{\textsc{IoT-NIDD}\xspace}

\newcommand{\ciciomtshort}{\textsc{IoMT}\xspace}
\newcommand{\edgeshort}{\textsc{Edge}\xspace}
\newcommand{\iotniddshort}{\textsc{NIDD}\xspace}

\newcommand{\ece}{\textsc{ECE}\xspace}

\newcommand{\fscore}[0]{F$1$-score\xspace}

\newcommand{\pauc}[0]{pAUC\xspace}

\newcommand{\fpr}[0]{FPR\xspace}

\newcommand{\ie}[0]{i.e.,\xspace}
\newcommand{\viz}[0]{viz.,\xspace}

\newcommand{\eg}[0]{e.g.,\xspace}

\newcommand{\ienum}[1]{\begin{enumerate*}[label=(\textit{\roman*})] #1 \end{enumerate*}}

\newcommand{\ange}[0]{\textsc{ANG$e$}\xspace}
\newcommand{\ampe}[0]{\textsc{AMP$e$}\xspace}

\newcommand{\qonly}[0]{\textsc{Dense}\xspace}

\definecolor{groupA}{RGB}{235, 247, 255}  
\definecolor{groupB}{RGB}{200, 240, 255}  
\definecolor{groupC}{RGB}{209, 255, 189} 
\definecolor{hybridColor}{RGB}{255, 245, 230}  

\definecolor{lightrow}{RGB}{240, 240, 240}
\definecolor{darkrow}{RGB}{240, 240, 240}

\usepackage[framemethod=tikz]{mdframed}
\newmdenv[
  innerlinewidth=0.8pt,
  roundcorner=5pt,   linecolor=gray!70!black,   backgroundcolor=gray!5,   skipabove=8pt,
  innertopmargin=4pt,
  splittopskip=4pt,
  frametitle={\noindent \faStickyNote[regular]\;\textbf{\textsf{Takeaway:}}},
    frametitlerule=true,
  frametitlefont=\normalsize\bfseries\color{gray!70!black},
  frametitlebackgroundcolor=gray!20,
  font=\sffamily\small,   nobreak=true,
]{takehomebox}

\definecolor{quantum}{HTML}{A9C4EB}
\definecolor{deep}{HTML}{B9E0A5}

\newcommand{\hybrid}[0]{\textsc{CNN+LSTM}\xspace}
\newcommand{\ampehybrid}{\ampe~(\hybrid)\xspace}

\definecolor{dlcolor}{RGB}{60, 180, 90}
\definecolor{qmlcolor}{RGB}{45, 120, 215}

\definecolor{dlcolor}{RGB}{103,156,235}

\newcommand{\fakefract}[0]{\texttt{FF}\xspace}
\newcommand{\fakelagos}[0]{\texttt{FLa}\xspace}
\newcommand{\fakelondon}[0]{\texttt{FLo}\xspace}
\newcommand{\fakenairobi}[0]{\texttt{FN}\xspace}

\newcommand{\fakeyorktown}[0]{\texttt{FY}\xspace}
\newcommand{\qdefault}[0]{\textit{Noiseless}\xspace}

\newcommand{\diamondicon}{\includegraphics[height=2.5ex]{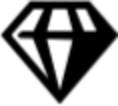}}

\newcommand{\backbone}[0]{Backbone\xspace}
\newcommand{\qcircuit}[0]{Quantum Circuit\xspace}
\newcommand{\cllayer}[0]{Classification Layer\xspace}

%% file: tex/00_abstract.tex
The rapid growth of the \gls{iot} has increased networked systems’ exposure to cyber threats, making \glspl{nids} essential for securing heterogeneous environments.
While \gls{dl} has shown strong performance in traffic analysis, \gls{qml} has recently gained attention as an alternative approach to learn complex traffic patterns.
We assess \gls{qml}’s practical applicability to intrusion detection in heterogeneous \gls{iot} scenarios.
We present a hybrid quantum–classical \gls{nids} and evaluate it on three recent \gls{iot} security datasets,
\emph{under operationally relevant conditions}.
Beyond detection accuracy, we compare \gls{qml} with a classical  \gls{dl} baseline across \emph{five operational perspectives}:
\ienum{\item performance under data scarcity, \item transferability through cross-domain evaluation, \item resilience to temporal constraints, \item robustness to quantum non-idealities across multiple noisy backends, and \item prediction trustworthiness through calibration analysis.
}
Experiments on binary and multiclass misuse detection show that \gls{dl} generally outperforms \gls{qml} under ideal and noisy conditions. However, \gls{qml} remains competitive in data-scarcity and domain-shift scenarios and can outperform the \gls{dl} baseline under temporal constraints.

\begin{IEEEkeywords}
Deep Learning,
Network Intrusion Detection System,
Quantum Machine Learning, Trustworthy AI.
\end{IEEEkeywords}

%% file: tex/010_introduction.tex
\glsresetall

The widespread adoption of the \gls{iot} enables a broad range of services such as remote monitoring, smart infrastructures, and safety-critical applications~\cite{zhou2017security}.
The increasing number of interconnected devices, including wearables and smart home systems, has significantly expanded both the volume and heterogeneity of network traffic.\footnote{\url{https://www.ericsson.com/en/reports-and-papers/mobility-report/reports/november-2025}}
However, the highly distributed and heterogeneous nature of \gls{iot} environments also enlarges the attack surface available to malicious actors. Cyberattacks targeting \gls{iot} systems may lead to severe consequences, including privacy breaches, financial losses, service disruption, and threats to human safety~\cite{bugshan2023intrusion}.
Consequently, significant research efforts have focused on developing solutions that detect cyber threats in a timely and reliable manner.
Among them, \glspl{nids} are designed to distinguish legitimate network activities from malicious traffic and identify the specific type of attack carried out by the adversary.
However, traditional approaches often struggle to cope with the increasing complexity, scale, and encryption of modern network traffic.
\gls{ml} and, more recently, \gls{dl} techniques have become central to
modern \glspl{nids}, especially in \gls{iot} scenarios, where the proliferation of connected devices considerably expands the attack surface.
\gls{qc} represents a promising frontier in information science, emerging as a powerful response to the continuous growth of \gls{iot} traffic data volumes, with
the ability to provide theoretical speedups on large-data processing and more accurate computational results.
The intersection of \gls{qc} paradigm with \gls{ml} has given rise to \gls{qml}.
\gls{qml} combines quantum mechanics principles with \glsentryshort{ai}-driven approaches~\cite{alchieri2021introduction, spadari2025towards}, offering a new way to capture non-linearity phenomena in complex data and a potential solution for pervasive security in \gls{iot} environments.
Despite the surging interest in this field, practical quantum advantage remains an open issue~\cite{spadari2025towards}, with empirical performance gains mostly limited to small-scale evaluations on synthetic or outdated datasets.
Inherent quantum hardware noise affects circuit measurements and, consequently, \gls{qml} outputs~\cite{tudisco2024evaluating, cirillo2026benchmarking}.
\gls{qml}-based \glspl{nids} also face training data scarcity, domain transferability challenges, and timing constraints.
Lastly, it remains unclear whether the confidence scores associated with classification decisions accurately reflect the actual predictive performance~\cite{nascita2024survey}.

\noindent
In
our investigation,
we consider the following
\emph{\glspl{oc}} for early network intrusion detection:
\begin{itemize}
    \item
            \textbf{Data Efficiency (\gls{oc}1):} \emph{{How does the \gls{nids} perform binary and multiclass misuse detection with scarce data?}}
    \item \textbf{Cross-Domain Generalization (\gls{oc}2):}
        \emph{Is the \gls{nids} transferable to domains different than its original one?}
    \item \textbf{Resilience to Temporal Constraints
    (\gls{oc}3):} \emph{{How robust is \gls{nids} detection under strict temporal constraints?}}
        \item \textbf{Robustness to Noise and Sampling Errors (\gls{oc}4):} \emph{How do physical hardware noise and finite-shot sampling affect \gls{qml}-based \gls{nids} performance?}
    \item \textbf{Trustworthiness Assessment (\gls{oc}5):} \emph{How reliable and well-calibrated are the \gls{nids} predictions?}
\end{itemize}

To address these challenges, we first discuss \gls{qml} for network intrusion detection and position our contribution with respect to prior work.
Then, we present the proposed hybrid quantum–classical \gls{nids} and the experimental methodology.
Finally, we evaluate the framework through the five \glspl{oc} to identify practical \gls{qml} opportunities and limitations for \gls{iot} network intrusion detection.
Data and code to reproduce all results are available at \url{https://github.com/viciEnzoh/qml-nids}.

%% file: tex/020_related_work.tex
Detecting attacks in network traffic is critical in \gls{iot} environments, where malicious activities can cause data leakage, service disruption, and economic losses.
Traditional \gls{ml} and \gls{dl} approaches have proven highly effective in addressing cybersecurity tasks~\cite{bugshan2023intrusion}.
However, their deployment in real-world \gls{iot} environments remains challenging due to severe resource constraints and the growing scale and heterogeneity of network traffic.
These limitations have motivated \gls{qc}-based alternatives, with \gls{qml} emerging as a promising direction for network security~\cite{biamonte2017quantum}. 
Recent studies have explored \gls{qml} for attack detection and classification~\cite{abreu2024qml, spadari2025towards, ahmad2026quantum}, but several operational challenges remain poorly addressed. 
First, most existing approaches rely on statistical (post-mortem) features extracted from complete traffic aggregates (\eg flows or biflows). 
Consequently, attacks are often detected only after communication has ended, limiting the \gls{nids} ability to react promptly and mitigate potential damage. 
The behavior of \gls{qml} models under limited training data also remains largely unexplored, despite being a common condition in cybersecurity, especially for rare or zero-day attacks. 
Generalization of \gls{qml} models remains under-researched as well.
Addressing this gap is key to a ``one \gls{nids} to detect them all'' paradigm~\cite{ismail2025intrusion, apruzzese2022}, where a single framework generalizes across heterogeneous network environments without retraining. 
Another critical challenge is the resilience to temporal constraints. 
In operational settings, reducing detection latency implies temporally-constrained traffic visibility, forcing \glspl{nids} to decide before attack signatures fully manifest.
This trade-off between responsiveness and detection accuracy has been studied for classical \gls{ml}/\gls{dl} models but has received limited attention in \gls{qml}. 
Beyond these operational challenges, deploying \gls{qml} in real-world network security environments is still limited by current quantum hardware.
While deployment is restricted by the limited availability and scale of actual quantum computers, current \gls{nisq} devices also suffer from noise and qubit decoherence, which can corrupt measurements and degrade \gls{nids} effectiveness. 
Recent studies using simulated and real quantum hardware~\cite{tudisco2024evaluating,cirillo2026benchmarking} provide limited insights into how hardware-induced noise affects model reliability. 
Hence, the trustworthiness of \gls{qml}-based \glspl{nids} under noisy hardware
remains an open challenge.

\vspace{3pt}
\noindent
\textbf{Positioning.}
This work extends our preliminary study~\cite{spadari2025towards}, which investigated the \emph{early detection}
of a \gls{qml}-powered \gls{nids} under fixed packet visibility
across
two \gls{iot}
datasets.
Here, we broaden the evaluation to \emph{three}
\gls{iot}
datasets spanning medical, industrial, and smart home domains under different \emph{data-scarcity} regimes.
We investigate key operational aspects previously overlooked.
We assess the impact of limited traffic observation time on the \gls{nids} performance via a maximum \emph{observation window}.
Then, we evaluate how \emph{quantum hardware noise} and \emph{finite-shot sampling} affect detection performance across multiple noisy quantum backends.
Finally, we assess \gls{nids} \emph{trustworthiness} via calibration analysis to determine whether predictive confidence aligns with classification accuracy, thereby validating the system’s operational reliability.

%% file: tex/030_methodology.tex
\input{floats/end2end}

The following describes the proposed hybrid classical--quantum pipeline for \emph{early} \gls{nids}, focusing on the quantum components of the architecture (\emph{Path B} in
Fig.~\ref{fig:end2end}).

\vspace{3pt}
\noindent
\textbf{Hybrid classical/quantum \gls{nids} pipeline.}
Network traffic is segmented into bidirectional flows (biflows) based on the standard five-tuple.
The \gls{nids} input is derived from header fields extracted from the first packets of each biflow, \ie without inspecting payloads.
The \gls{nids} performs \emph{early detection}, enabling the identification of malicious activity before the attack is fully established.
To optimize response time,
biflows can be processed in parallel across both classical and quantum layers, provided sufficient resources, \viz qubit counts, are available.
Biflows are used to feed a hybrid \emph{\gls{dl}/\gls{qml}-based model}.
Structurally, the model pipeline is composed as follows:
\ienum{
\item a classical \emph{\backbone}
compressing data
to satisfy quantum input constraints,
\item a \emph{\qcircuit} processing the compressed data, and
\item a final \emph{\cllayer} providing the output label.}
The framework can scale to
\ienum{
\item higher-dimensional features via backbone compression tailored to
feature-map requirements, and \item larger
deployments by extending the classification layer
for a wider attack spectrum.}

\vspace{3pt}
\noindent
\textbf{Attack Detection and Classification via
\glspl{qnn}.}
\gls{qc} relies on qubits, through properties like superposition and entanglement, for optimized information encoding.
To achieve this, \gls{qml} effectively exploits quantum mechanics principles by employing parameterized quantum circuits in complex input processing.
In this context, \gls{qml} employs parameterized circuits to model complex non-linearities by projecting classical data into higher-dimensional Hilbert spaces~\cite{biamonte2017quantum, alchieri2021introduction}.
\glspl{qnn} represent the most common approach~\cite{reejisha2023improving, ding2025qfmlnet},
thanks to their reduced architectural complexity and low parameter footprint.
As shown in the \qcircuit block of Fig.~\ref{fig:end2end}, these networks are based on three fundamental elements:
\ienum{
\item a feature map,
\item an ansatz, and
\item a measurement process.
}
The \emph{feature map}
translates
classical data into a quantum representation (quantum encoding).
In this work, we consider two popular
strategies:
\ienum{\item
Amplitude embedding (\ampe),
that encodes a normalized feature vector
into the probability amplitudes of a quantum state
offering a highly compact data representation;
\item
Angle embedding (\ange),
that converts
features into rotation angles,
requiring one qubit per field.}
The \emph{ansatz} is the trainable component of the quantum layer.
It
consists of sequential layers composed of rotation gates,
whose angles represent the weights,
adjusted during
optimization.
Each layer consists of single- and multi-qubit gates.
Quantum processing outputs are converted back into a classical representation through a \emph{measurement process}.
In this work, the circuit produces a probability vector of
size $2^{N_q}$, where $N_q$ is the number of qubits, with
each entry representing the probability of measuring a specific computational basis state.
This joint representation captures both individual qubit behavior and entanglement-induced correlations.
Crucially, under noisy conditions, empirical probability estimation via finite measurements ($S$ \emph{shots}) affects overall runtime latency rather than structural circuit complexity.
Finally, the probability vector feeds a \emph{classification} stage, where softmax produces class probabilities, and the most likely label identifies benign traffic or a specific attack.
\emph{Training} follows classical gradient-based optimization, minimizing a loss function between the predicted soft output and the one-hot encoded ground truth.

%% file: floats/end2end.tex
\begin{figure*}

    \centering
    \includegraphics[trim=0 0 0 0, clip, width=.95\textwidth]{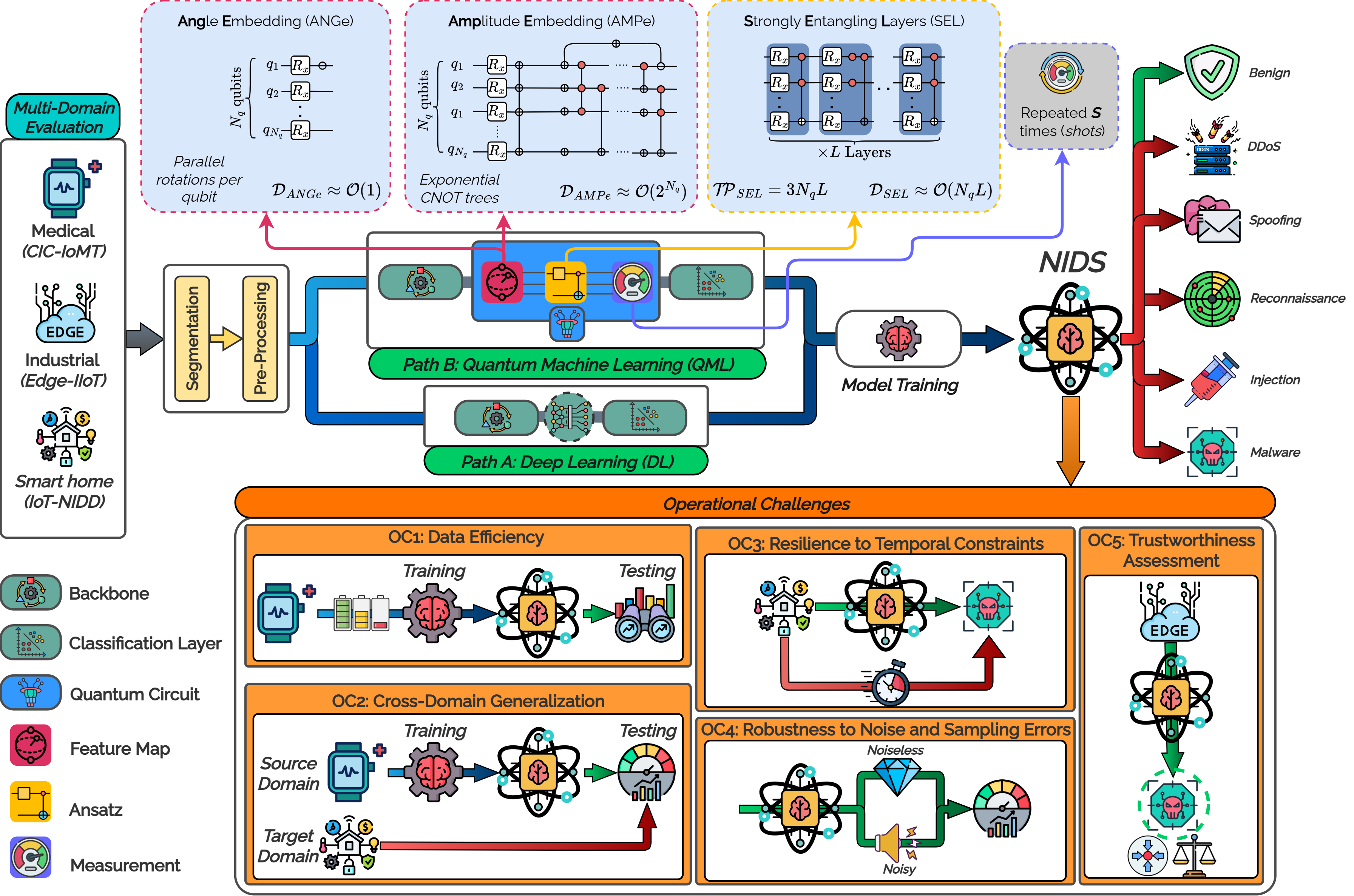}
    \caption{
    End-to-end hybrid
        pipeline for multi-domain \gls{iot} \gls{nids}.
        Each domain is represented by a distinct dataset. After traffic segmentation and pre-processing, the pipeline splits into two paths: (A)
        \gls{dl} and (B) \gls{qml}, the latter featuring a classical backbone-based dimensionality reduction, a quantum variational circuit, and a classical classification head.
    The quantum variational circuit consists of: ($i$) a feature map (\ampe, \ange), ($ii$) an ansatz (\gls{sel}), and ($iii$) a measurement process, executed over a fixed number ($S$) of shots.
    The framework maps network traffic to threat intelligence, distinguishing benign from attack traffic (\eg
    DDos, Malware), while addressing five \emph{Operational Challenges (OCs)}: (\gls{oc}1) data efficiency, (\gls{oc}2) cross-domain generalization, (\gls{oc}3) resilience to temporal constraints,
    (\gls{oc}4) robustness to noise and sampling errors,
    and (\gls{oc}5) trustworthiness assessment.
    Circuit complexity is characterized in terms of circuit depth ($\mathcal{D}$) and number of trainable parameters ($\mathcal{TP}$).}
    \label{fig:end2end}
\end{figure*}

%% file: tex/050_experimental_results.tex
\noindent
\textbf{Datasets.}
To ensure a comprehensive evaluation, our analysis spans three distinct \gls{iot} domains: medical, industrial, and
smart home.
This
allows us to characterize the specific threat landscape of each domain and to practically assess cross-domain \gls{nids} portability.
Accordingly, we leverage \emph{three}
\gls{iot} datasets, each corresponding to one of the aforementioned domains: \ciciomt (\ciciomtshort), \edge (\edgeshort), and \iotnidd (\iotniddshort).
These datasets
encompass benign traffic, and
$16$, $13$, and $9$ distinct attack types across approximately $604$k, $731$k, and $201$k biflows, respectively, covering a broad spectrum of network threats, \eg, (D)DoS,
Man-in-the-Middle, Injection,
and Malware.
For each biflow, we extract $4$ header fields from the first
packets:
\gls{dir}, \gls{ps}, \gls{tcpwin}, and \gls{iat}.
We consider
$10$ packets, as they provide a suitable trade-off between earliness and detection capabilities.

\vspace{3pt}
\noindent
\textbf{Proposed Models.}
We
compare the performance of a classical \gls{dl} architecture against various \gls{qml} configurations.
As a robust baseline for
binary and multiclass misuse detection,
we consider \hybrid~\cite{lopez2017network}, which consists of alternating convolutional and pooling layers followed by an LSTM layer.
The \gls{qml} models
leverage \ampe and \ange feature maps,
preceded by an initial dense layer (\qonly) for data compression.
For the variational component, we adopt the \glsentrylong{sel} ansatz.
Furthermore,
since \ampe was found to be the most effective feature map in our evaluation, we introduce \ampehybrid, an extended model that integrates \gls{dl}-based data compression by leveraging a \hybrid backbone before the \ampe stage.
In all hybrid models, the backbone acts as an independent dimensionality-reduction stage, decoupling the classical pipeline from the downstream quantum circuit.
Notably, \hybrid and the deep compression of \ampehybrid have the same structure but different bottleneck size, \ie, $100$ and $32$, respectively.
Hence, they represent two complete model configurations, whose parameters are optimized independently during training.
In line with qubit counts in the state of the art~\cite{cirillo2026benchmarking, ahmad2026quantum},
all simulations employ $5$ qubits, due to \gls{nisq}-era constraints.

\vspace{3pt}
\noindent
\textbf{Quantum Devices Simulation.}
The models are implemented
with the PennyLane simulation device \texttt{default.qubit}.
Initial training is conducted in \emph{analytic mode} to establish performance under ideal ({\emph{Noiseless}})
conditions.
However, to account for the inherent noise, decoherence, and gate errors characterizing \gls{nisq} devices, we extend the evaluation to noisy scenarios.
We leverage the \texttt{qiskit.aer} simulator to reproduce the noise profiles of real IBM hardware (\emph{fake backends}).
Specifically, these noisy profiles represent static snapshots of real quantum processors, capturing the target's physical error rates, qubit decoherence, and readout infidelity at a given point in their lifetime.
Given a fake backend,
expectation values are estimated from a finite number of
\emph{shots} ($S$),
approximating the probability distribution of the circuit's basis states to account for operational hardware constraints.
The analytic mode adopted for the \emph{Noiseless} reference corresponds to
$S\!=\!\infty$.
Thus, deviations from it jointly reflect
\ienum{
\item finite-shot measurement error, and
\item simulated quantum device noise.
}

\vspace{3pt}
\noindent
\textbf{Evaluation Strategy.}
Evaluation follows the well-established stratified $5$-fold cross-validation protocol.
For each fold, we reserve $20\%$ of the training data for validation and apply \emph{MinMax} normalization fitted only on the training set to prevent data leakage.
Experiments are conducted on both
\emph{binary} and \emph{multi-class} misuse detection scenarios, leveraging the \gls{pauc} and (macro) \fscore as primary metrics, respectively.
The \gls{pauc} measures the (partial) area under the \gls{roc} curve for a fixed \gls{fpr} interval (\gls{pauc} at $1\%$\gls{fpr} in our setup). We prioritize a low \fpr threshold to meet operational requirements for network service continuity in real-world \gls{nids} deployments, thus limiting false alarms.
Conversely, the (macro) \fscore averages the harmonic means of per-class precision and recall, providing a balanced performance measure across both benign traffic and each specific attack class.
It quantifies the \gls{nids} capability not only to isolate benign from malicious traffic, but also to differentiate among attack types.
Finally, we assess calibration in the multi-class scenario via the \gls{ece}, which quantifies the deviation between predicted confidence and actual accuracy using $10$ bins over the maximum predicted class probabilities.
A higher \gls{ece} indicates poorer calibration and lower reliability of the \gls{nids} confidence estimates
when distinguishing between traffic classes.

\section*{Operationalizing Quantum NIDS: \\Five Critical Perspectives}

\input{floats/results/01_f1_vs_pauc_scatter}

\noindent
\textbf{\gls{oc}1:
Data Efficiency.}
We first analyze the ability of models to generalize under limited training data. This setting reflects real-world \gls{iot} security scenarios, where emerging threats and zero-day attacks often lack sufficient samples for training \glspl{nids}.
To emulate these scenarios, we systematically vary the training set fraction
$\delta$
and evaluate the resulting impact on performance.
We vary $\delta\!\in\!\{10, 50, 100\}\%$, spanning one order of magnitude while retaining an adequate number of training samples. Fig.~\ref{fig:f1_vs_pauc_performance} compares the binary (\gls{pauc} at $1\%$ \gls{fpr}) with the multi-class misuse detection performance (\fscore) for each \gls{qml} model as the training set size increases.
For comparison, we also include the results of the \gls{dl} baseline (\hybrid).
Our analysis reveals \emph{three} primary performance drivers that dictate the effectiveness of the proposed architectures: (\textit{i}) the expressivity of the feature map, (\textit{ii}) the depth of the data-compression backbone, and (\textit{iii}) the impact of the quantum processing.
Among the \emph{quantum encoding} strategies, \ange is the least effective, particularly on \ciciomtshort and \iotniddshort. On the former, for instance, performance does not exceed $54\%$ \fscore and $61\%$ \pauc, highlighting its limited capacity to capture complex threat patterns in these environments.
In contrast, \ampe demonstrates superior representative power, achieving a \fscore (\pauc) of up to $69\%$ ($95\%$) on the same dataset.
This encoding also highlights remarkable data efficiency:
training on only $50\%$ of the dataset
yields performance within $3\%$ of the model trained on the entire dataset, suggesting that quantum embedding captures the threat signatures with significantly fewer samples.

The \emph{data-compression} performed by the backbone is equally decisive.
In fact, by holding the \ampe feature map constant, we observe that transitioning from a shallow architecture (\qonly) to a deeper one (\hybrid) triggers a significant performance boost, even under extreme data scarcity ($\delta\!=\!10\%$), where the \pauc remains above $91\%$ on \ciciomtshort.
Interestingly, on \edgeshort, performance saturates at $50\%$ training data, obtaining $>\!99.8\%$ of \pauc.
This suggests that a reduced, well-stratified training set acts as an implicit regularizer, enabling the model to ignore stochastic network fluctuations and prioritize extracting universal threat signatures.

Finally, the performance gap between \hybrid and \ampehybrid, which are powered by the same
backbone architecture,
is at most $3\%$ in \fscore and $1\%$ in \pauc in full data regimes, highlighting the dominant role of the common compression structure.
Still, as the two models do not share the parameters adjusted during training, this gap
\emph{is not fully motivated by the quantum circuit alone}.
However, a notable disparity emerges under extreme data scarcity: at $\delta\!=\!10\%$, the \ampehybrid architecture exhibits a significantly lower \fscore than its classical counterpart,
with a gap of $\approx\!15\%$ on the \iotniddshort dataset.
This indicates that the quantum circuit requires more samples to achieve the same level of training stability as classical layers.
Based on these findings, since \ampehybrid demonstrates the highest overall effectiveness, we select it for subsequent in-depth analyses against the \hybrid \gls{dl} baseline.

\input{floats/results/02_crosseval_tpr}
\vspace{3pt}
\noindent
\textbf{\gls{oc}2: Cross-domain Generalization.} We conduct a cross-domain evaluation to assess the robustness of \gls{dl} and \gls{qml} models under \emph{domain shift}.
The goal is to determine whether the models can generalize to unfamiliar network environments without requiring site-specific retraining.
This analysis is particularly important for practical \gls{nids} deployment, since models trained and tested on the same network data distribution may unintentionally hide environment-specific patterns that limit their portability.
In our evaluation, we train a model on a \emph{source} dataset and evaluate it using benign biflows from the \emph{source} domain and malicious biflows from a \emph{target} domain (source$\rightarrow$target).
This strategy allows us to assess whether attack patterns learned on the source network
enable the detection of malicious traffic originating from different networks.
Moreover, keeping benign samples from the source domain in both training and testing helps reduce false alarms after deployment~\cite{apruzzese2022}.
Fig.~\ref{fig:crosseval-pauc} shows the \gls{pauc} at $1\%$ \gls{fpr} for all
source$\rightarrow$target
pairs, comparing \ampehybrid and \hybrid.
The \emph{in-domain} baseline is shown on the main diagonal, with both benign and malicious samples from the same dataset.
As expected, in-domain performance is consistently high, with both models achieving a \pauc exceeding $95\%$ across all datasets. This confirms that the architectures are highly effective when the training and testing distributions align.
However, robustness to domain shift varies significantly across source datasets. Models trained on \edgeshort demonstrate superior portability, maintaining a \pauc of at least $67\%$ across all unfamiliar environments.
In contrast, training on \iotniddshort leads to significant, and in some cases catastrophic, performance degradation ($\text{\pauc}\!<\!17\%$), with models struggling to distinguish benign from malicious traffic. This lack of generalization is likely due to the specific composition of \iotniddshort, where approximately $80\%$ of the biflows consist of a single packet.
Such a limited observation window precludes the learning of complex temporal or structural patterns, forcing models to overfit to environment-specific artifacts that do not translate into portable attack representations.
Comparing \ampehybrid and \hybrid reveals considerable discrepancies in cross-domain generalization between the two configurations.
In
\ciciomtshort{}$\rightarrow$\iotniddshort, \ampehybrid achieves $+23\%$ average \pauc compared to the \gls{dl} baseline,
yet it shows a $6\%$ decrease in \edgeshort{}$\rightarrow$\ciciomtshort,
suggesting that the quantum-enhanced feature mapping could extract more resilient cross-domain representations.
However,
the large standard deviations across folds make it challenging to identify a clear winner.
Rather, a trend can be inferred
from fold-by-fold inspection.
Accordingly,
Fig.~\ref{fig:crosseval-pauc} also highlights cases where one model achieves higher \pauc in at least $4$ out of $5$ folds across several configurations.
These results underscore that cross-domain robustness remains a critical, yet frequently overlooked, challenge.

\input{floats/results/03_tti}

\vspace{3pt}
\noindent
\textbf{\gls{oc}3: Resilience to Temporal Constraints.}
We conduct a fine-grained analysis to assess the capability of the \gls{nids} to provide accurate detection \emph{under limited traffic observation time}, a critical factor for rapid threat mitigation and damage minimization to network equipment or users under attack.
To simulate time-bounded observation, we enforce a maximum observation window $\tau$ for each biflow, whereby only packets observed within this window are used for inference.
This operational setup differs from \emph{earliness} since the latter restricts the analysis to a fixed packet count ($N_p$), whereas $\tau$ imposes a hard temporal constraint.
Consequently, if the first $N_p$ packets are not fully collected before $\tau$ expires, classification is triggered using only the available packets at that instant.
Fig.~\ref{fig:tti-f1} depicts the \fscore achieved by \hybrid and \ampehybrid models on all datasets as $\tau$ increases. We also include the ``unconstrained'' scenario ($\tau\!=\!\infty$).
The experimental results across different datasets reveal distinct behaviors regarding the sensitivity to the observation window $\tau$.
On \edgeshort, we observe a marginal performance gap of approximately $2\%$ between \ampehybrid and \hybrid when enforcing strict operational latency constraints ($\tau\!\leq\!1$s). This discrepancy progressively narrows as $\tau$ increases, with both architectures performing similarly as early as $\tau\!=\!10$s.
Conversely, performance on \iotniddshort remains largely invariant to the observation window. In fact, both models exhibit near-identical behavior regardless of $\tau$, a result that aligns with our previous observation regarding the dataset's composition: since the majority of biflows are extremely short, enforcing a temporal deadline has a negligible impact on the available information for detection.
The most significant divergence occurs on
\ciciomtshort.
In the high-response regime ($\tau\!\leq\!100$ms), \ampehybrid achieves a higher \fscore than \hybrid, suggesting that quantum-enhanced feature mapping might be more effective at extracting threat signatures under limited traffic visibility.
However, this trend reverses for $\tau\!\geq\!500$ms, where the baseline emerges as the top-performing model.
Notably, both models show increased instability for
$\tau\!\in\![0.5, 10]$s,
as evidenced by the wider variance.
This suggests that these time intervals capture a transition phase in which the available packets may not yet represent a stable fingerprint of the underlying threat. Lastly, when no responsiveness constraints are enforced on the system, the performance gap remains consistent, with \hybrid outperforming \ampehybrid by $\approx\!3\%$ \fscore.

\input{floats/results/04_noisy_backends_f1}
\vspace{3pt}
\noindent
\textbf{\gls{oc}4: Robustness to Noise and Sampling Errors.}
Contemporary quantum hardware is inherently susceptible to environmental decoherence and gate-level errors, which act as signal distorters in a \gls{nids} pipeline.
To quantify the impact of physical noise {and finite-shot sampling} on detection capability, we compare the ideal simulation  (\qdefault) against \emph{five} noisy backends: \emph{FakeLagos} (\fakelagos), \emph{FakeNairobi} (\fakenairobi), \emph{FakeLondon} (\fakelondon), \emph{FakeFractional} (\fakefract), and \emph{FakeYorktown} (\fakeyorktown).
These backends are selected to span a wide spectrum of noise profiles, capturing the evolution of \gls{nisq} technologies
from legacy platforms (\eg \fakeyorktown) to mid-generation architectures (\eg \fakelagos and \fakenairobi).
This heterogeneity allows us to evaluate the framework under diverse physical conditions, including readout errors arising from imperfect qubit measurement and limited qubit coherence times, \ie the duration a qubit retains its quantum state before undergoing environmental decoherence.
Crucially, featuring up to $7$ qubits,
they
align with our feature mapping dimensionality, ensuring a realistic assessment of physical \gls{nisq} constraints.
Fig.~\ref{fig:noisy-pauc} depicts the \fscore of \ampehybrid across all datasets.
We assess stochastic sampling by varying
$S\!\in\!\{8, 64, 1024\}$
to evaluate the trade-off between measurement overhead and detection accuracy, identifying the minimum sampling depth required for satisfactory operational performance.
As expected,
$S$
is a primary performance driver, with scaling from $8$ to $1024$
generally
yielding a substantial improvement in \fscore.
A notable exception occurs on \iotniddshort, where \fakelagos and \fakeyorktown show a decreasing \fscore as $S$ increases.
A possible interpretation is that,
under high-noise conditions,
\eg high readout errors or short coherence times, increasing $S$
merely consolidates systematic hardware bias, potentially degrading performance.
On the one hand, operating with only $8$ shots is practically infeasible across all backends, with a gap of $\ge\!10\%$ (\eg \fakenairobi on \ciciomtshort) compared to \qdefault,
due to the combined effect of low-shot sampling variance and hardware noise.
On the other hand, \fakenairobi and \fakelondon emerge as the most noise-resilient architectures.
This likely stems from their
efficient topologies (H- and T-shape) and modern native gates  (\texttt{cx}/\texttt{sx}),
yielding near-optimal performance with just $64$ shots and approaching the baseline
across almost all datasets (\eg losing at most $2\%$ with \fakenairobi on \ciciomtshort and \edgeshort).

\input{tables/tab_ece_fbs_signed} \vspace{3pt}
\noindent
\textbf{\gls{oc}5: Trustworthiness Assessment.} To assess the predictive reliability of the proposed pipeline under realistic deployments, we conduct a calibration analysis. While model calibration is paramount in classical \glspl{nids}~\cite{nascita2024survey}, evaluating it in the \gls{nisq} era is critical. In fact, hardware-induced noise, systematic biases, and gate inaccuracies directly corrupt the output probability distributions of quantum circuits, potentially leading to misaligned predictive confidence. Consequently, a model might achieve high accuracy but exhibit a severe discrepancy between empirical performance and predicted confidence. This mismatch manifests as a high \gls{ece}, indicating that the quantum \glspl{nids} are miscalibrated.
From an operational standpoint, hardware-induced miscalibration may lead to catastrophic security failures, such as generating false alarms or leaving actual intrusions unnoticed.
Conversely, a well-calibrated model ensures that output probabilities accurately reflect the true likelihood of its predictions. Hence, \gls{ece} assesses whether advances in modern quantum backends translate into reliable operational benefits by reducing alarm overload through effective thresholding strategies.
Table~\ref{tab:ece_shots} reports the \gls{ece} achieved by \ampehybrid across all datasets for each (Backend, Shots) pair. For comparison, we also report performance for the \qdefault scenario
and the \emph{\gls{dl} baseline}.
Since both models share the same backbone architecture, this comparison serves a twofold purpose:
\ienum{
\item evaluating the
effect of the hybrid classical-quantum design on model calibration and
\item assessing how calibration changes under noisy finite-shot execution compared to the analytic \emph{Noiseless} setting.
}
On the one hand, adopting a hybrid model does not degrade calibration.
Instead, it yields slight improvements on \edgeshort and \iotniddshort, allowing both models to exhibit excellent calibration on all datasets with an \gls{ece} consistently below $1\%$.
On the other hand, comparing the backends against the \qdefault scenario confirms the trends observed in our previous analysis, corroborating that noisy finite-shot execution could be the cause of degraded calibration.
Specifically, the \gls{ece} under realistic hardware environments remains consistently higher than in the {\emph{Noiseless}} baseline.
Moreover, calibration instability is further substantiated by the high variability across some noisy backends, particularly \fakelagos and \fakeyorktown, which demonstrate relevant standard deviations in their \gls{ece}.
Among the evaluated backends, \fakenairobi is the only one in which the \gls{ece} monotonically improves with increasing shots, crossing the sub-$1\%$ \gls{ece} threshold exclusively on
\edgeshort.
An inspection of the reliability diagrams, omitted for brevity, reveals that the quantum circuit tends to reduce predictive confidence, leading to under-confidence. Although this helps maintain calibration under noiseless conditions, hardware noise exacerbates the effect, degrading output probabilities and causing misclassifications.
These findings suggest that the noiseless regime targeted by future fault-tolerant quantum architectures enables well-calibrated \gls{qml}-based \glspl{nids}, whereas current quantum hardware suffers calibration degradation from hardware noise and finite-shot uncertainty.

%% file: floats/results/01_f1_vs_pauc_scatter.tex
\begin{figure}[t]
    \centering

    \subfloat{
        \includegraphics[trim=0 450 0 0, clip, width=.86\columnwidth]{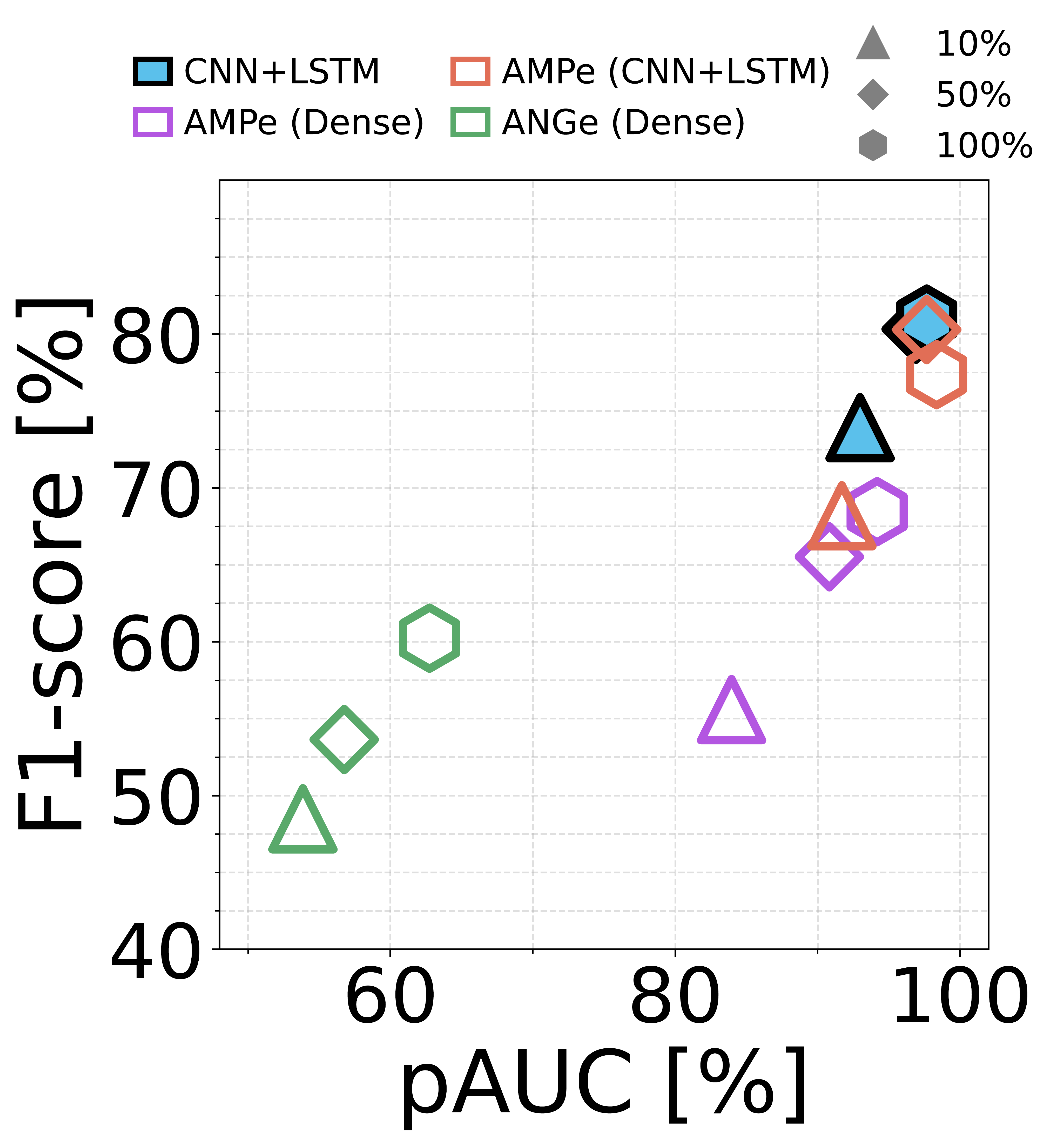}
    }
        \vspace{-10pt}
    \setcounter{subfigure}{0}
    \subfloat[\ciciomtshort\label{fig:ciciomt-f1-vs-pauc}]{
        \includegraphics[trim=0 0 0 80, clip, height=.3\columnwidth]{img/01_f1_vs_pauc/cic-iomt_f1_vs_pauc2.png}
    }
    \subfloat[\edgeshort\label{fig:edgeiiot-f1-vs-pauc}]{
        \includegraphics[trim=50 0 0 80, clip, height=.3\columnwidth]{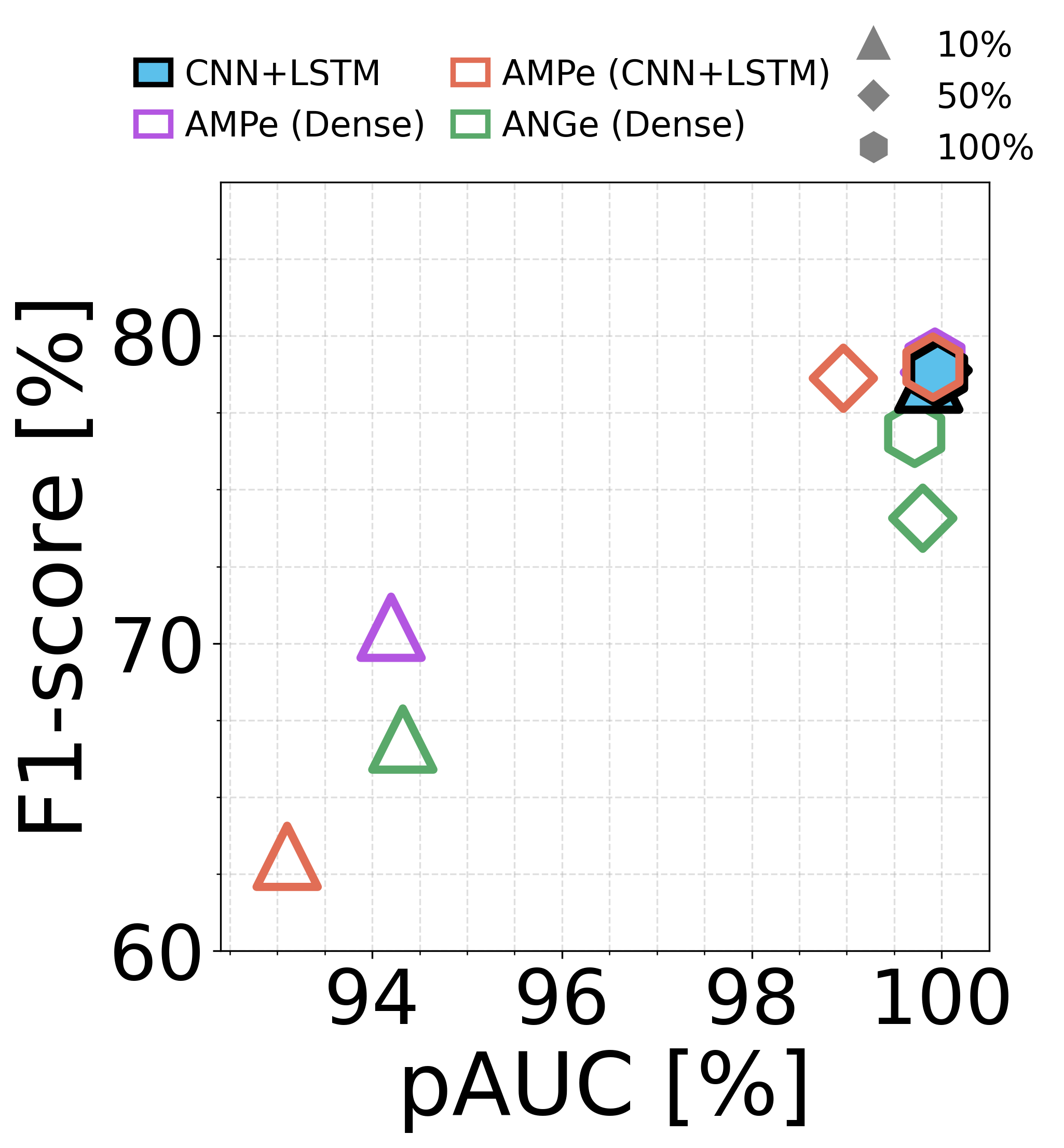}
    }
    \subfloat[\iotniddshort\label{fig:iotnidd-f1-vs-pauc}]{
        \includegraphics[trim=50 0 0 80, clip, height=.3\columnwidth]{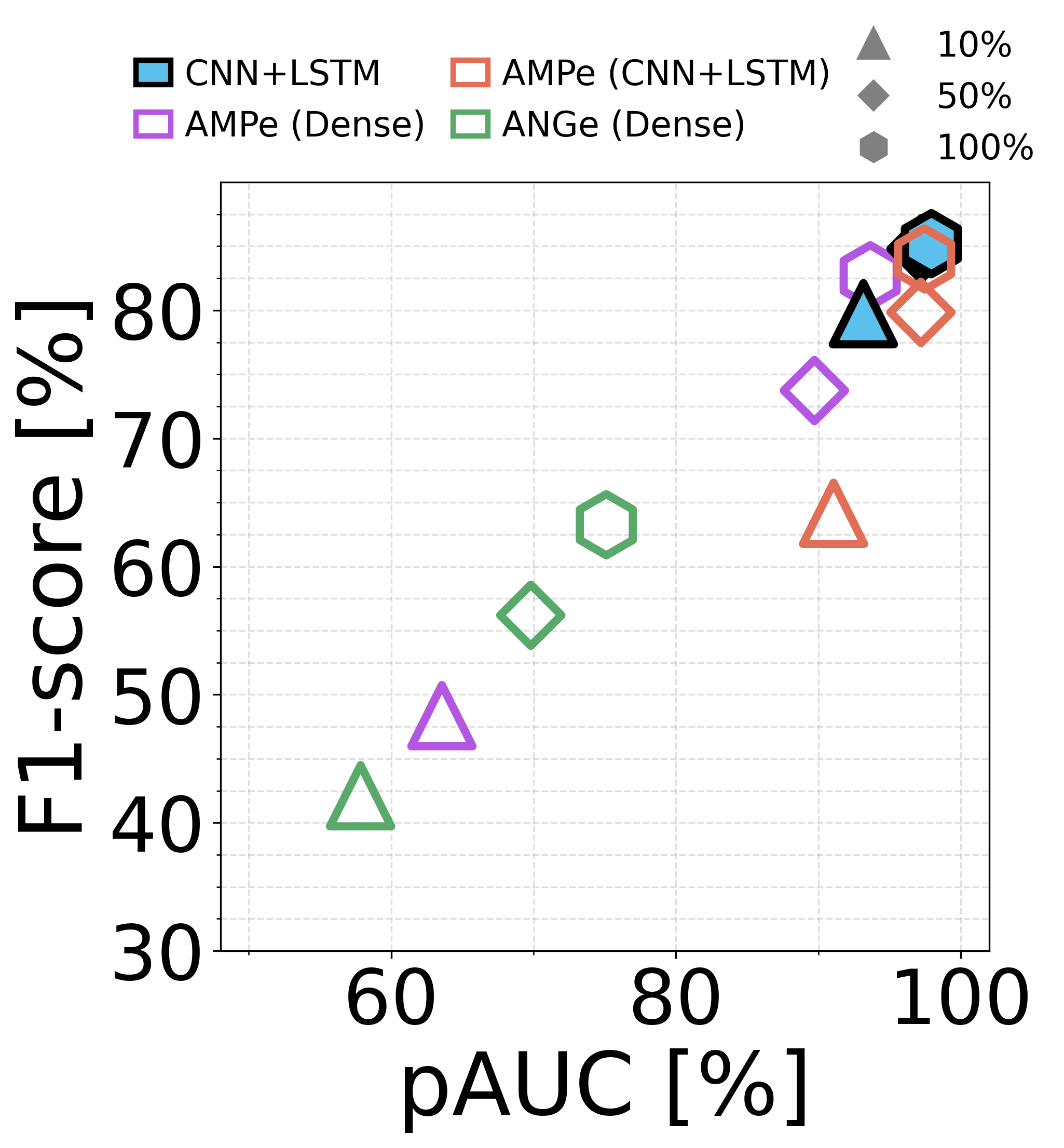}
    }
            \caption{
        \fscore vs. \pauc (\fpr max. $1\%$) for \gls{dl} and \gls{qml} models on
        all datasets across distinct training data proportions ($\delta\!\in\!\{10, 50, 100\}\%$).
        Values are averaged across $5$ folds.
                }
    \label{fig:f1_vs_pauc_performance}
\end{figure}

%% file: floats/results/02_crosseval_tpr.tex
\begin{figure}[t]

\vspace{-12pt}
    \centering
    \includegraphics[trim=40 0 130 32, clip, height=0.35\columnwidth]{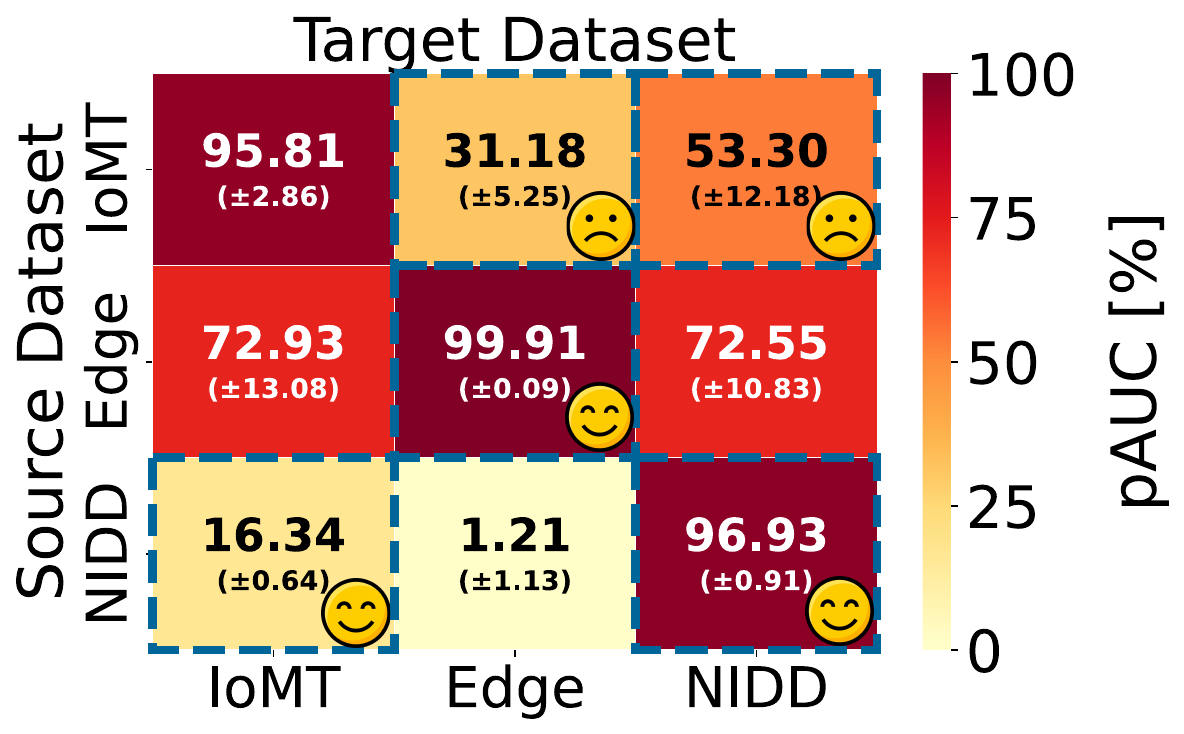}
    \includegraphics[trim=70 0 130 32, clip, height=0.35\columnwidth]{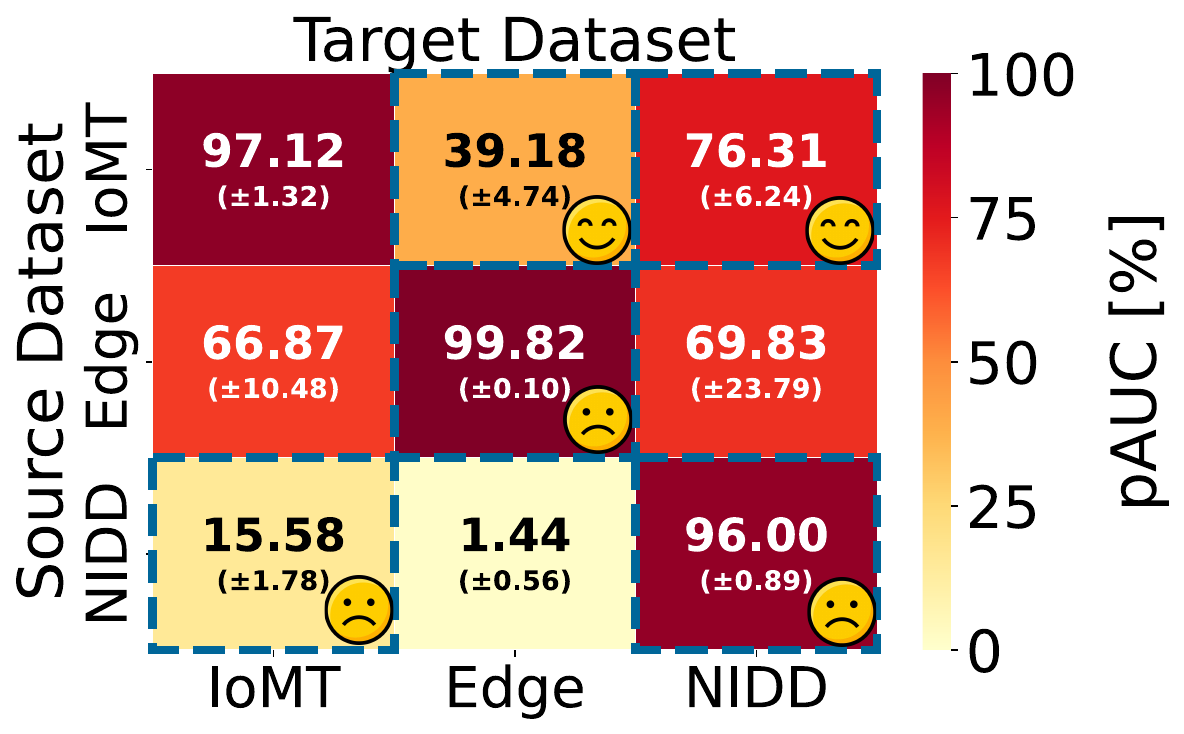}
        \subfloat{
        \includegraphics[trim=430 0 50 0, clip, height=.38\columnwidth]{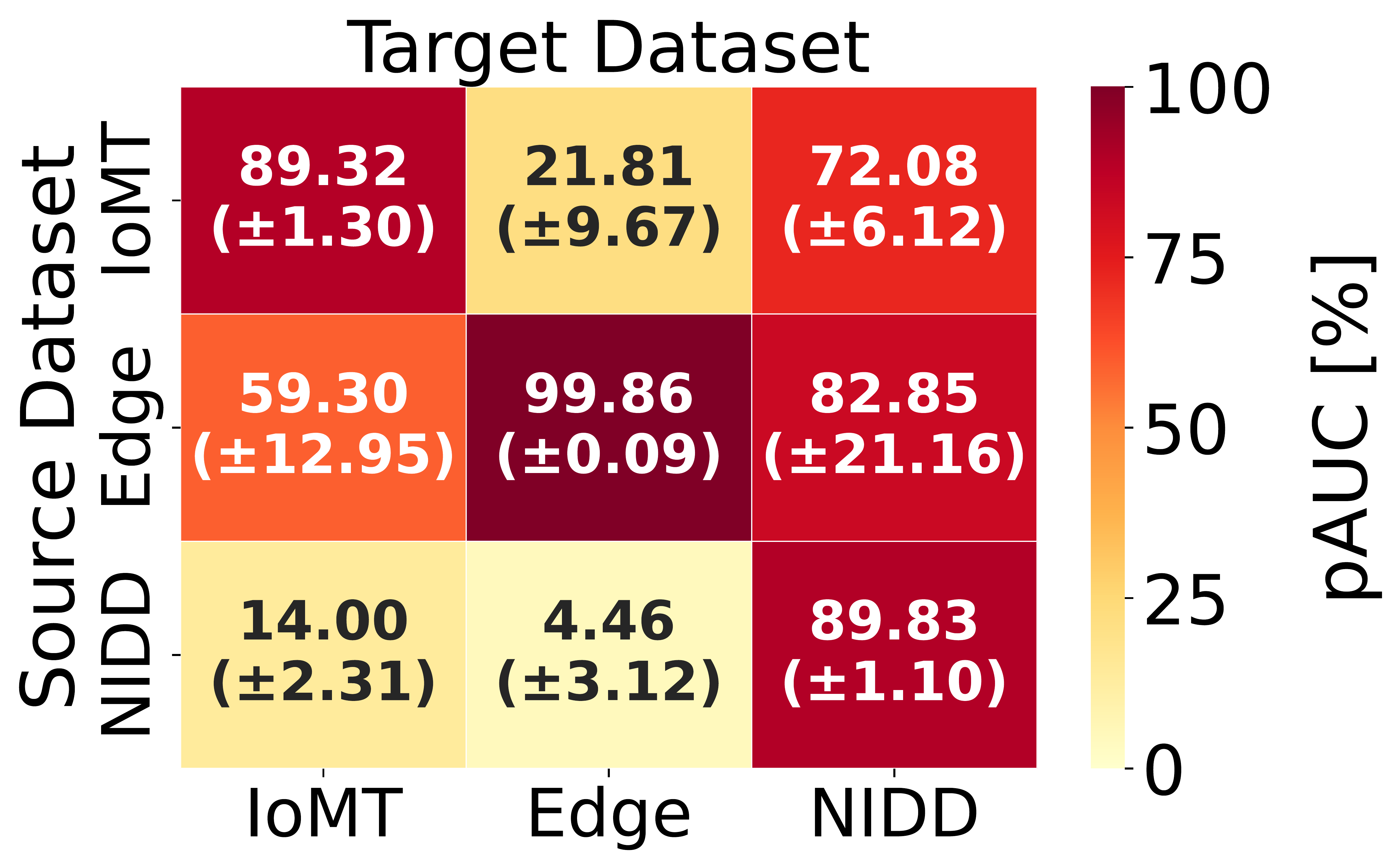}
    }
    \caption{
                            \pauc (\fpr max. $1\%$) of \hybrid (\textbf{left}) and \ampehybrid (\textbf{right}) in cross-evaluation setup. Rows/columns denote source/target datasets, respectively. Values are $mean \pm std.dev.$ across $5$ folds.
        {
    Dashed boxes highlight configurations where         \faSmile[regular] achieves higher \pauc than \faFrown[regular] in $\ge 4/5$ folds.}
        }
    \label{fig:crosseval-pauc}
\end{figure}

%% file: floats/results/03_tti.tex
\begin{figure}[t]
    \centering
    \includegraphics[trim=0 0 0 0, clip, width=0.99\columnwidth]{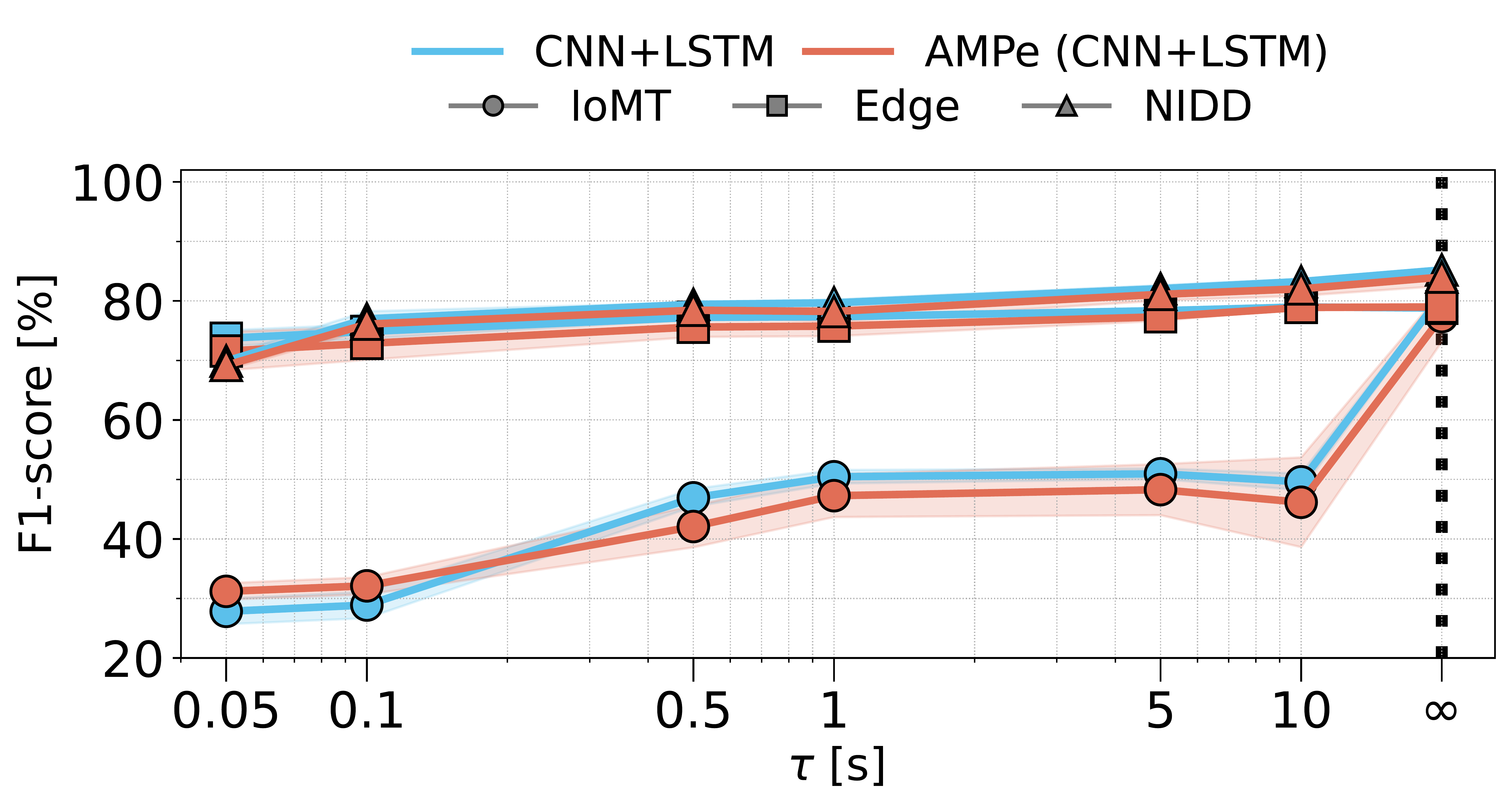}
    \caption{
        \fscore of \ampehybrid and \hybrid as the maximum observation window ($\tau$) increases. $\tau\!=\!\infty$: unconstrained scenario.
    Values are $avg.\pm std.$ across $5$ folds.
        $x$-axis is in \textit{log} scale.
        }
    \label{fig:tti-f1}
\end{figure}

%% file: floats/results/04_noisy_backends_f1.tex
\begin{figure}[t]
    \centering
    \includegraphics[trim=0 0 5 0, clip, width=0.99\columnwidth]{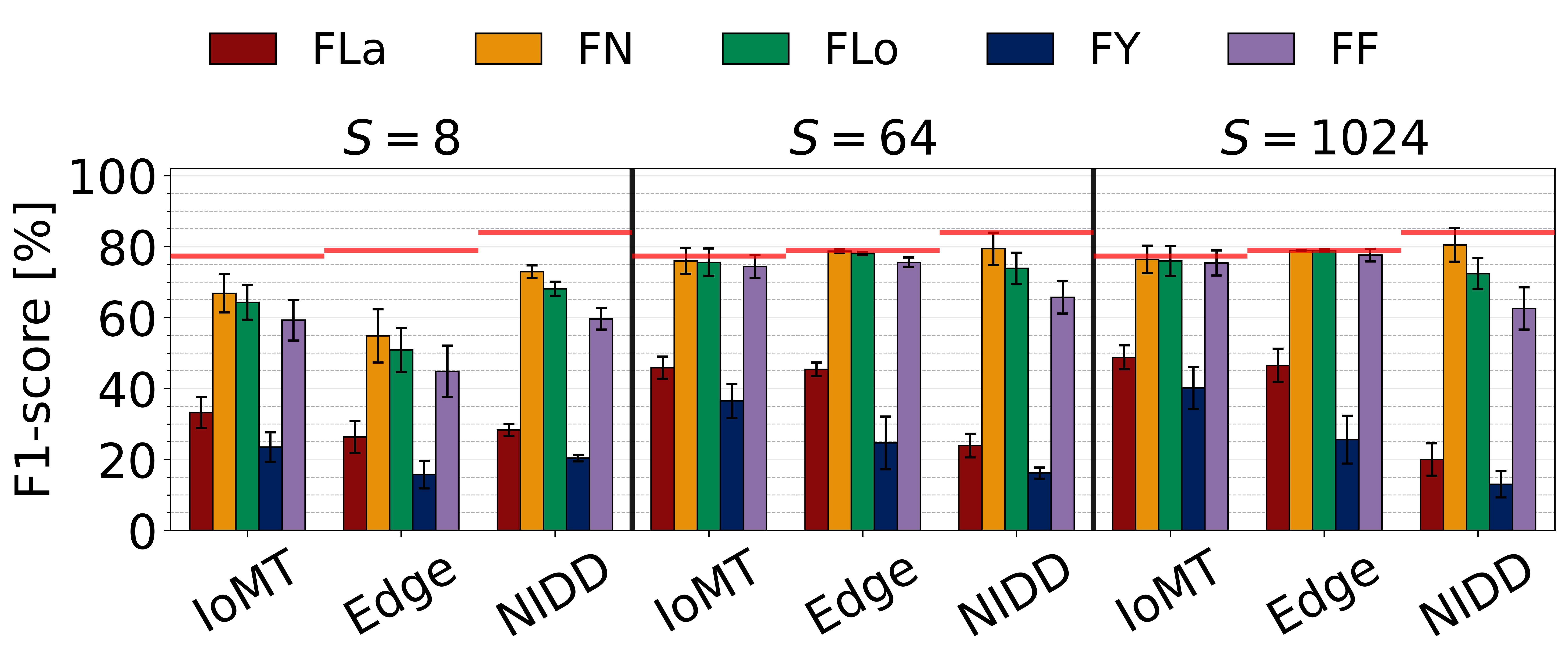}
    \caption{\fscore of
    \ampehybrid for different backends and shots ($S$).
    \textcolor{red}{\textbf{---}}: {\emph{Noiseless}} backend.
    Values are $avg.\pm std.$ across $5$ folds.
        }
    \label{fig:noisy-pauc}
\end{figure}

%% file: tables/tab_ece_fbs_signed.tex
\newcommand{\tinytiny}[1]{\fontsize{5}{10}\selectfont $\,\pm\,#1$}

\definecolor{pastelgreen}{HTML}{C8E6C9}
\definecolor{pastelred}{HTML}{FFCDD2}

\begin{table}[t]
  \centering
  \caption{
  \scriptsize
  \ece of \ampehybrid.
  Values are averaged across $5$ folds.
  Green (resp. red) cells highlight under- (resp. over-)confident models.
  \\\faDatabase: Dataset. \textbf{S}: Shots. \diamondicon: Noiseless. \faFlagCheckered: \gls{dl} baseline.    }
  \label{tab:ece_shots}
    \includegraphics[width=0.95\columnwidth]{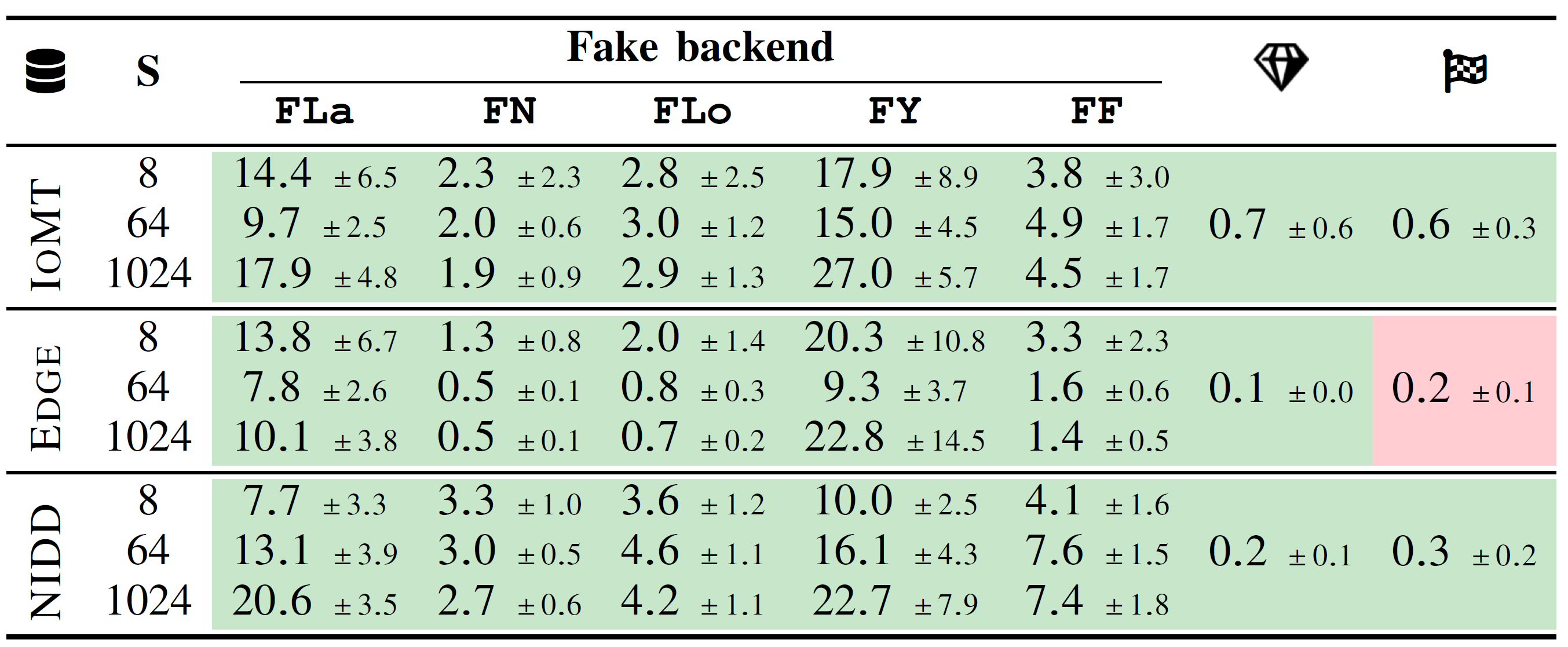}
\end{table}

%% file: tex/060_conclusion.tex
This work provides a practical assessment of a \gls{qml}-based \gls{nids} for early intrusion detection under \emph{operationally relevant conditions},
spanning data efficiency, cross-domain generalization, temporal resilience, robustness to noise and sampling errors, and prediction trustworthiness.
In several scenarios, \gls{qml} remains competitive
under limited training data (\textbf{\gls{oc}1}), suggesting their potential applicability to emerging threats and zero-day attacks.
The results highlight the importance of classical compression for high-quality quantum embedding and circuit optimization, helping narrow the performance gap between current \gls{qml} and \gls{dl}.
At the same time, portability across heterogeneous network environments remains a major challenge (\textbf{\gls{oc}2}). Cross-domain performance depends on the characteristics of the training dataset, emphasizing the difficulty of learning attack representations that generalize beyond environment-specific traffic patterns, for both classical and quantum models.
The
analysis of resilience to temporal
constraints
(\textbf{\gls{oc}3}) shows that \gls{qml} can improve early-stage intrusion detection, particularly when operating with limited traffic information.
Our study highlights that current limitations of quantum hardware cannot be overlooked in operational deployments (\textbf{\gls{oc}4}). Detection performance and model calibration are sensitive to hardware-induced noise, qubit decoherence, and backend-specific characteristics. While recent noisy backends approach noiseless behavior under favorable operating conditions, some hardware configurations still induce miscalibrated predictions, potentially compromising the reliability of generated alarms (\textbf{\gls{oc}5}).
Overall, \gls{qml} emerges as a viable component of future intelligent cybersecurity systems, particularly for early and adaptive intrusion detection in heterogeneous \gls{iot} environments.
From a deployment perspective, cloud queuing and orchestration overheads can hinder inline inspection, whereas compact few-qubit architectures could facilitate on-premises implementation.

\noindent
\textbf{Future work}
will explore larger quantum architectures with more qubits, advanced feature maps, ansatze, measurement schemes, and bottleneck-matched ablations to isolate quantum contributions.
We will extend OC2 to a full deployment setting, drawing benign and attack traffic from the target network,
to assess false alarms and portability.
Moreover, noiseless simulations with matched shot budgets will decouple finite-shot sampling error from physical noise.
We plan physical quantum hardware deployments to evaluate end-to-end decision latency and throughput beyond simulations,
with post-hoc calibration to mitigate sampling- and noise-related miscalibration.
Finally, fault-tolerant regimes will be investigated to understand \gls{oc} behavior when hardware noise becomes less relevant,
while finite-sampling and networking constraints remain.

%% file: tex/90_biographies.tex
\vskip -2.5\baselineskip plus -1fil
\begin{IEEEbiographynophoto}{Domenico Ciuonzo} [SM] (domenico.ciuonzo@unina.it) is an Associate Professor at the University of Napoli Federico II. His research concerns data fusion, network analytics, IoT, signal processing, and AI.
\end{IEEEbiographynophoto}

\vskip -2.5\baselineskip plus -1fil
\begin{IEEEbiographynophoto}{Idio Guarino} [M] (idio.guarino@unibo.it)
is an Assistant Professor at the University of Bologna, Italy. His research focuses on Internet traffic analytics, developing data-driven methods to model, classify, and predict network traffic.
\end{IEEEbiographynophoto}

\vskip -2.5\baselineskip plus -1fil
\begin{IEEEbiographynophoto}{Alfredo Nascita} (alfredo.nascita@unina.it) is an Assistant Professor at the University of Napoli Federico II. His research interests include Internet
traffic analysis, machine and deep learning, and explainable artificial intelligence.
\end{IEEEbiographynophoto}

\vskip -2.5\baselineskip plus -1fil
\begin{IEEEbiographynophoto}{Antonio Pescapè} [SM] (pescape@unina.it) is a Full Professor at the University of Napoli Federico II. His work focuses on measurement, monitoring, and analysis of the Internet.
\end{IEEEbiographynophoto}

\vskip -2.5\baselineskip plus -1fil
\begin{IEEEbiographynophoto}{Vincenzo Spadari} (vincenzo.spadari@unina.it) is a Ph.D Student at the University of Napoli Federico II. His research focuses on the application of quantum machine learning on network traffic data analysis.
\end{IEEEbiographynophoto}